\documentclass[times]{speauth}
\usepackage{subfig}
\usepackage{moreverb}
\usepackage{graphicx}
\usepackage{multirow}

\usepackage{amsmath}
\usepackage{amsfonts}
\usepackage[ruled, norelsize]{algorithm2e}
 \usepackage{caption}
\usepackage{amssymb,url,amsmath}

\newcommand\BibTeX{{\rmfamily B\kern-.05em \textsc{i\kern-.025em b}\kern-.08em
		T\kern-.1667em\lower.7ex\hbox{E}\kern-.125emX}}

\def\volumeyear{2016}

\begin{document}
	
\runningheads{H. Gupta, A.V. Dastjerdi, S.K. Ghosh, R. Buyya}{iFogSim: A Toolkit for Modeling and Simulation of Internet of Things}

\title{iFogSim: A Toolkit for Modeling and Simulation of Resource Management Techniques in Internet of Things, Edge and Fog Computing Environments}

\author{Harshit Gupta \footnotemark[1] \footnotemark[2], Amir Vahid Dastjerdi \footnotemark[1], Soumya K. Ghosh\footnotemark[2], and Rajkumar Buyya \footnotemark[1]}

\footnotetext[1]{H. Gupta, A.V Dastjerdi, and R Buyya are with Cloud Computing and Distributed Systems (CLOUDS) Laboratory,
Department of Computing and Information Systems, The University of Melbourne, Australia.\\Email: \{amir.vahid,rbuyya\}@unimelb.edu.au,  
}
\footnotetext[2]{
 H. Gupta and S.K. Ghosh are with the Department of Computer Science and Engineering, Indian Institute of Technology Kharagpur, India. \\ Email: harshitgupta1337@gmail.com, skg@iitkgp.ac.in
}
  
	\begin{abstract}
Internet of Things (IoT) aims to bring every object (e.g. smart cameras, wearable, environmental sensors, home appliances, and vehicles) online, hence generating massive amounts of data that can overwhelm storage systems and data analytics applications. Cloud computing offers services at the infrastructure level that can scale to IoT storage and processing requirements. However, there are applications such as health monitoring and emergency response  that require low latency, and delay caused  by transferring data to the cloud and then back to the application can seriously impact their performances. To overcome this limitation, Fog computing paradigm has been proposed, where cloud services are extended to the edge of the network to decrease the latency and network congestion. To realize the full potential of Fog and IoT paradigms for real-time analytics, several challenges need to be addressed. 
The first and most critical problem is designing resource management techniques that determine which modules of analytics applications are pushed to each edge device to minimize the latency and maximize the throughput. To this end, we need a evaluation platform that enables the quantification of performance of resource management policies on an IoT or Fog computing infrastructure in a repeatable manner. In this paper we propose  a simulator, called iFogSim, to model IoT and Fog environments and measure the impact of resource management techniques in terms of latency, network congestion, energy consumption, and cost. 
We describe two case studies to demonstrate modeling of an IoT environment and comparison of resource management policies. 
Moreover, scalability of the simulation toolkit in terms of RAM consumption and execution time is verified under different circumstances.	
	\end{abstract}
	
	
	\maketitle
		
	\vspace{-6pt}
	
	\section{Introduction}
	
The Internet of Things (IoT) paradigm promises to make "things" including consumer electronic devices or home appliances such as medical devices, fridge, cameras, and sensors part of the Internet environment. This paradigm opens  the  doors  to  new innovations  that  will  build  novel  type  of  interactions  among  things  and  humans  and enables the realization of smart cities, infrastructures, and services for enhancing the quality of life and utilization of resources. It supports integration, transfer, and analytics of data generated by smart devices (e.g. sensors). IoT envisions a new world of connected devices and humans in which quality of life is enhanced, because management of city and its infrastructure is less cumbersome, health services are conveniently accessible, and disaster recovery is more efficient. Based on bottom-up analysis for IoT applications, McKinsey estimates that the IoT has a potential economic impact of 11 trillion dollar per year by 2025--- which would be equivalent to about 11 percent of the world economy.  They also expect one trillion IoT devices will be deployed by 2025.

\begin{figure*}[htb!]
\centering
\includegraphics[width = .9\textwidth]{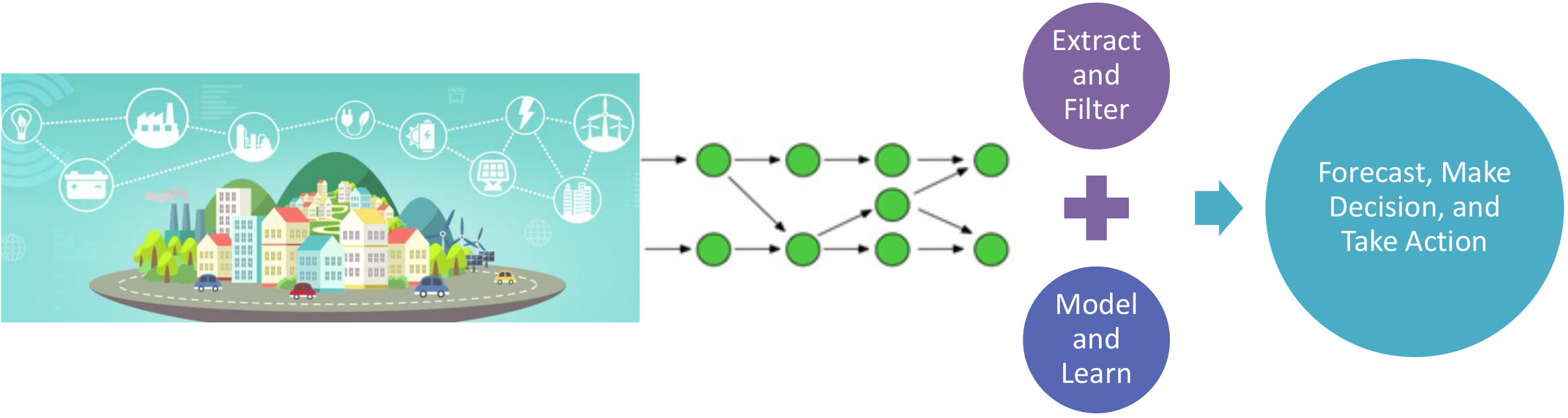}
\caption{IoT Environment Common Objectives.}
\label{fig:objective}
\end{figure*}

 Although technologies and solutions enabling connectivity and data delivery are growing rapidly, not enough attention has been given to real-time analytics and decision making as one of the major objectives of IoT (Figure \ref{fig:objective}).
Majority of current IoT data processing solutions transfer the data to cloud for processing. This is mainly because existing data analytics approaches are designed to deal with large volume of data, but not real-time data processing and dispatching. With millions of \textit{things} generating data, transferring all of that to the cloud is neither scalable nor suitable for real-time decision making. The dynamic nature of IoT environments and its associated real-time requirements and increasing processing capacity of edge devices (entry point into provider core networks, e.g. gateways) \cite{chang2015middleware} has lead to the evolution of the Fog computing paradigm. Fog computing \cite{bonomi2014fog} extends cloud services to the edge of networks, which results in latency reduction through geographical distribution of IoT application components, and provides support for mobile mobility. 

Many IoT applications (e.g.  stream processing) are naturally distributed and are often embedded in an environment with numerous connected computing devices with heterogeneous capabilities. As data travels from its point of origin (e.g. sensors) towards applications deployed in Cloud virtual machines, it passes through many devices, each of which is a potential target of computation offloading. Therefore, it is important to take advantage of computational and storage capabilities of these intermediate devices. 
The main challenge lies in scheduling application components (operators in terms of stream processing or independent short-lived IoT tasks) in the pool of devices --- from the network edge to the cloud --- to meet application level Quality-of-Service (QoS) requirements such as end-to-end latency or privacy requirements while minimizing resource and energy wastage. 

Resource management policies to ensure Quality of Service (QoS), avoid energy wastage, and resource fragmentation are an integral part of IoT systems. To foster innovation and development enabling real-time analytics in Fog computing, we require a an evaluation environment for exploring different resource management and scheduling techniques including operator (operator and application module have been used interchangeably in the paper) and task placement, migration and consolidation.  A real IoT environment as a testbed, although  desirable, in many cases is too costly and does not provide repeatable and controllable environment. To address this shortcoming, we propose a simulator called iFogSim that enables the simulation of resource management and application scheduling policies across edge and cloud resources under different scenarios and conditions. 

In this paper, we discuss the architecture of iFogSim along with its design and implementation. The framework is designed in a way that makes it capable of evaluation of resource management policies applicable to Fog environments with respect to their impact on latency (timeliness), energy consumption, network congestion and operational costs. It simulates edge devices, cloud data centers, and network links to measure performance metrics. The major application model considered for iFogSim is the Sense-Process-Actuate model. In such models, sensors publish data to IoT networks, applications running on Fog devices subscribe to and process data coming from sensors, and finally insights obtained are translated to actions forwarded to actuators. In addition, we present a simple IoT simulation recipe and  two case studies to demonstrate how one can model an IoT environment and plug in and compare resource management policies. Finally, we evaluate the scalability of iFogSim in terms of memory consumption and simulation execution time.

\par The paper is structured as follows: A formal definition of Fog computing, its concepts and benefits are presented in Section \ref{sec:definition}. Section \ref{sec:architecture} discusses the architecture of iFogSim followed by its implementation details, sample resource management policies, and a generic simulation recipe in Section \ref{sec:implementation}. Case studies along with their application models and network topologies are expounded in Section \ref{sec:case_study}. Section \ref{sec:pe} presents results of scalability tests on iFogSim and compares two basic resource management policies considering metrics such as latency, energy consumption and network usage. Finally, Section \ref{sec:conclusion} concludes the paper and discusses the future directions.

\begin{figure*}[!b]
\centering
\includegraphics[width = .9\textwidth]{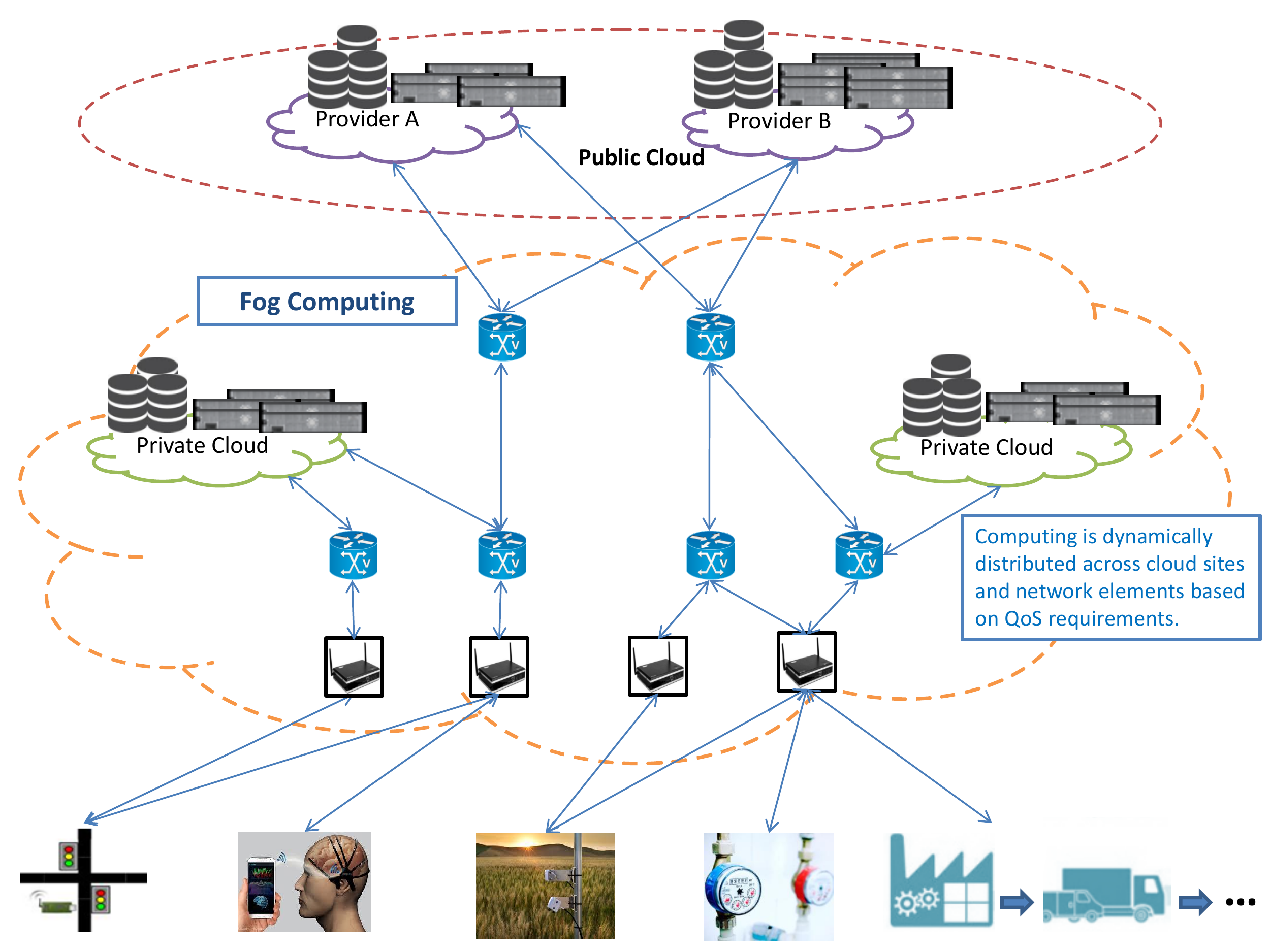}
\caption{Distributed Data Processing in a Fog Computing Environment.}
\label{fig:env}
\end{figure*}

\section{Fog Computing --- Definition and Concepts}
\label{sec:definition}
We define Fog computing as a distributed computing paradigm that extends the services provided by the cloud to the edge of the network. It enables seamless leveraging of cloud and edge resources along with its own infrastructure (see Figure \ref{fig:env}). It facilitates management and programming of compute, networking and storage services between data centers and end devices. Fog computing essentially involves components of an application running both in the cloud as well as in devices between endpoints and the cloud, i.e. smart gateways and routers. Fog computing supports mobility, resource and interface heterogeneity, interplay with the cloud, and distributed data analytics to addresses requirements of applications that need low latency with a wide and dense geographical distribution. Fog computing takes advantages of both edge and cloud computing ---  while it benefits from edge devices' close proximity to the endpoints, it also leverages the on-demand scalability of cloud resources.
\begin{figure*}[b!]
\centering
\includegraphics[width = .9\textwidth]{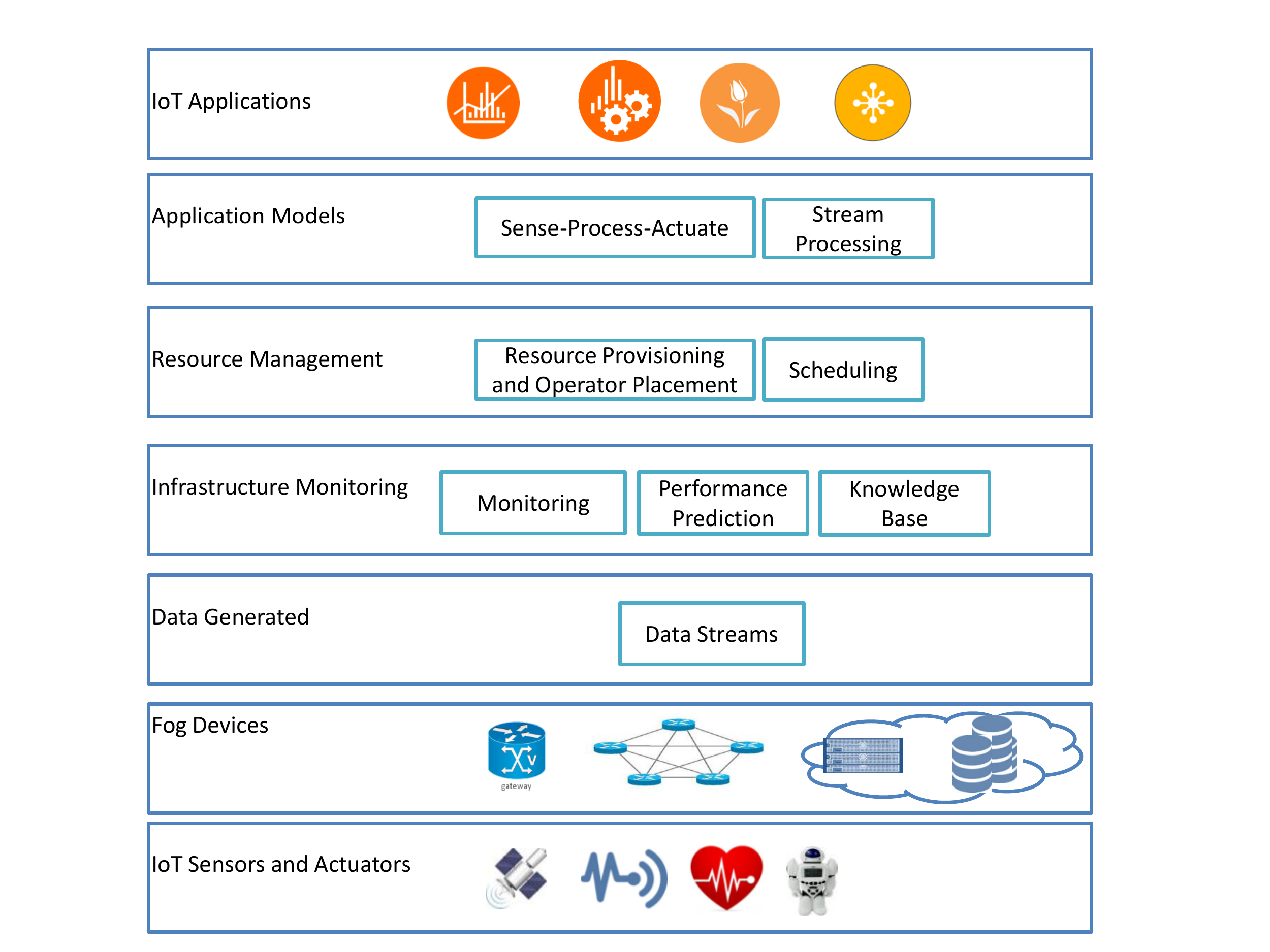}
\caption{Fog Computing Architecture.}
\label{fig:arch}
\end{figure*}
\par There are a number of benefits associated with Fog computing that assures its success. The first benefit is the reduction of network traffic as uncontrolled increase in network traffic may lead to congestion, and results in increased latency. 
Fog computing provides a platform for filtering and analysis of the data generated by sensors by utilizing resources of edge devices.
This drastically reduces the traffic being sent to the cloud by allowing the placement of filtering operators close to the source of data. Considerable reduction in propagation latency is the next important advantage of utilizing Fog computing paradigm especially for mission critical applications that require real-time data processing. Some of the best examples of such applications are cloud robotics, control of fly-by-wire aircraft, or anti-lock brakes on a vehicle. In the case of cloud robotics, the effectiveness of motion control is contingent on the timeliness of data collection by the sensors, processing of the control system and feedback to the actuators. Having the control system running on the cloud may make the sense-process-actuate loop slow or unavailable as a result of communication failures. This is where Fog computing helps by performing the processing of the  control system very close to the robots --- thus making real-time response possible. Finally, cloud computing paradigm, even with it's virtually infinite resources, can become a bottleneck if all the raw data generated by end devices (sensors) is sent to a centralized cloud. Fog computing is capable of filtering and processing considerable amount of incoming data on edge devices, making the data processing architecture distributed and thereby scalable.

\section{Architecture}
\label{sec:architecture}
 The architecture of Fog computing environment as presented in Figure \ref{fig:arch} involves a hierarchical arrangement of Fog nodes throughout the network between sensors and the cloud at the core of the network. In the architecture, \textbf{IoT sensors} are placed at the bottommost layer of the architecture and distributed in different geographical locations, sensing the environment, and emitting observed values to upper layers via gateways for further processing and filtering. Similarly, \textbf{IoT Actuators} operate at the bottommost layer of the architecture and are responsible for controlling a mechanism or system. Actuators are usually designed to respond to changes in environments that are captured by sensors. \textbf{IoT Data Streams} are made of a sequence of immutable values emitted by sensors. In the architecture any element in the network that is capable of hosting application modules is called \textbf{Fog Device}. Fog devices that connect sensors to the Internet are generally called gateways. Fog devices also include cloud resources that are provisioned on-demand from geographically distributed data centers. 
 
 In addition, the architecture defines three main services and two application models for Fog and IoT environments that are described below.
 
\textbf{Monitoring components} keep track of the resource utilization and availability of sensors, actuators, Fog devices and network elements. They keep track of the applications and services deployed on the infrastructure by monitoring their performance and status. Monitoring components supply this information to other services as required.

\textbf{Resource management} is the core component of the architecture and consists of components that coherently manage resources in such a way that application level QoS constraints are met and resource wastage is minimized. To this end, Placement and Scheduler components play a major role by keeping track of the state of available resources (information provided by the Monitoring service) to identify the best candidates for hosting an application module.

\textbf{Power monitoring:} 
One of the toughest challenges that most IoT solutions face is utilization of resources of IoT nodes while considering constraints on energy consumption. In contrast to cloud data centers, Fog computing encompasses a large number of devices with heterogeneous power consumption, making energy management difficult to achieve. 
Hence, evaluating the impact of applications and resource management policies on energy consumption is crucial before deployment in production environments. Therefore, we require a power monitoring component in the architecture that is responsible for monitoring and reporting the 
energy consumption of Fog devices in the simulation.

\textbf{Application (programming) models:} The applications developed for deployment in the Fog are based on the Distributed Data Flow (DDF) model \cite{giang2015developing}. An application is modelled as a collection of modules, which constitute the data processing elements. Data generated as output by module $i$ may be used as input by another module $j$, giving rise to data dependency between module $i$ and $j$. This application model allows us to represent an application in the form of a directed graph, with the vertices representing application modules and directed edges showing the flow of data between modules. Later we present two sample applications modeled as DDF. 


The architecture supports two models used for IoT applications:
\begin{enumerate}
\item Sense-Process-Actuate Model: The information collected by sensors is emitted as data streams, which is acted upon by applications running on Fog devices and the resultant commands are sent to actuators.
\item Stream Processing Model: The stream processing model has a network of application modules running on Fog devices that continuously process data streams emitted from sensors. The information mined from the incoming streams is stored in data centers for large-scale and long-term analytics.
\end{enumerate} 
We consider Stream Processing model as a subcategory of the Sense-Process-Actuate model. These models can, however, be extended to cater to use-cases other than IoT applications.

\section{Design and Implementation}
\label{sec:implementation}

\begin{figure*}[!htb]
\centering
\includegraphics[width = .90\textwidth]{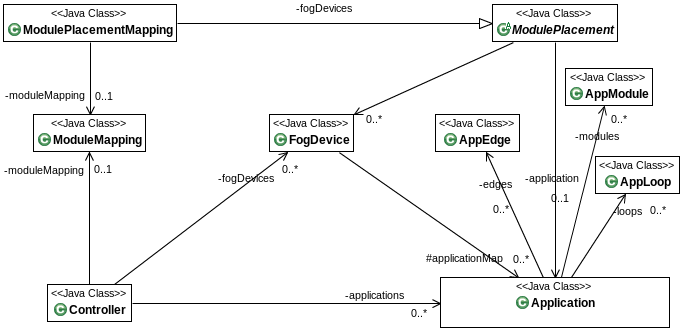}
\caption{Main classes of iFogSim.}
\label{fig:classDiagram}
\end{figure*}

\begin{figure*}[!b]
\centering
\includegraphics[width = .9\textwidth]{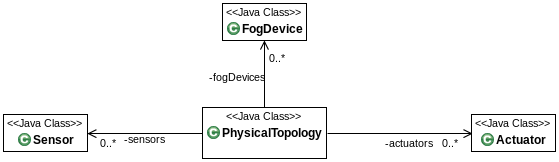}
\caption{iFogSim physical topology classes.}
\label{fig:classtopo}
\end{figure*}

For implementing functionalities of iFogSim architecture, we leveraged basic event simulation functionalities found in CloudSim\cite{calheiros2011cloudsim}. Entities in CloudSim, like data centers, communicate between each other by message passing operations (sending events, to be more precise). Hence, the core CloudSim layer is responsible for handling events between Fog computing components in \textit{iFogSim}. The main classes of \emph{iFogSim} are depicted in Figures~\ref{fig:classDiagram} and \ref{fig:classtopo}. In this section, we present the details of these classes and their interactions. The implementation of \textit{iFogSim} is constituted by simulated entities and services. First, we describe how the elements of architecture are modeled as iFogsim classes.				

\begin{itemize}
\item \textbf{FogDevice:} This class specifies hardware characteristics of Fog devices and their connections to other Fog devices, sensors and actuators. Having been realized by extension from the \textit{PowerDatacenter} class in CloudSim, the major attributes of the FogDevice class are accessible memory, processor, storage size, uplink and downlink bandwidths (defining the communication capacity of Fog devices). Methods in this class define how the resources of a Fog device are scheduled between application modules running on it and how modules are deployed and decommissioned on them. Overriding these methods enables developers to plug-in custom policies for the abovementioned functions.
\item \textbf{Sensor:} Instances of the Sensor class are entities that act as IoT sensors described in the architecture. The class contains attributes representing the characteristics of a sensor, ranging from its connectivity to output attributes. 
The class contains a reference attribute to the gateway Fog device to which the sensor is connected and the latency of connection between them. Most importantly, it defines the output characteristics of the sensor and the distribution of tuple inter-transmission or inter-arrival time - which identifies the tuple arrival rate at the gateway.

\item \textbf{Tuple:} Tuples form the fundamental unit of communication between entities in the Fog. Tuples are represented as instances of Tuple class in iFogSim, which is inherited from the \textit{Cloudlet} class of CloudSim. A tuple is characterized by its type and the source and destination application modules. The attributes of the class specify the processing requirements (defined as Million Instructions (MI)) and the length of data encapsulated in the tuple. 
\item \textbf{Actuator:} This class models an actuator by defining its network  connection properties. An attribute in the class refers to the gateway to which the actuator is connected and the latency of this connection. The class defines a method to perform an action on arrival of a tuple from an application module.
\item \textbf{Application:} 
An application is modeled as a directed graph, the vertices of the DAG representing modules that perform processing on incoming data and edges denoting data-dependencies between modules. These entities are realized using the following classes:

\begin{itemize}
	\item \textbf{AppModule}: Instances of \textit{AppModule} class represent processing elements of Fog applications. \textit{AppModule} is implemented by extending the class \textit{PowerVm} in CloudSim. For each incoming tuple, an \textit{AppModule} instance processes it and generates output tuples that are sent to next modules in the DAG. The number of output tuples per input tuple is decided using a selectivity model --- which can be based on a fractional selectivity or a bursty model.


   \item  \textbf{AppEdge}: An \textit{AppEdge} instance denotes the data-dependency between a pair of application modules and represents a directed edge in the application model. Each edge is characterized by the type of tuple it carries, which is captured by the \textit{tupleType} attribute of AppEdge class along with the processing requirements and length of data encapsulated in these tuples. iFogSim supports two types of application edges --- periodic and event-based. Tuples on a periodic \textit{AppEdge} are emitted at regular intervals. A tuple on an event-based edge $e=(u,v)$ is sent when the source module $u$ receives a tuple and the selectivity model of $u$ allows the emission of tuples carried by $e$.
   

	\item \textbf{AppLoop:} \textit{AppLoop} is an additional class, used for specifying the process-control loops of interest to the user. In iFogSim, the developer can specify the control loops to measure the end-to-end latency. An AppLoop instance is fundamentally a list of modules starting from the origin of the loop to the module where the loop terminates.
 \end{itemize}

\end{itemize}

\begin{figure*}[!htb]
\center
\includegraphics[width=.9\textwidth]{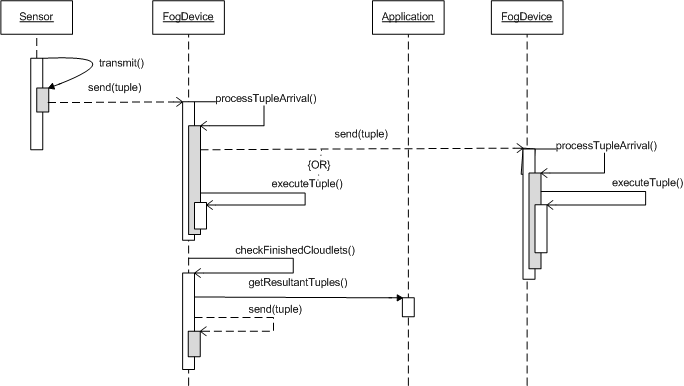}
\caption{Sequence diagram of the generation and execution.}
\label{sqdiagram}
\end{figure*}

A sequence diagram demonstrating tuple emission and subsequent execution is shown in Figure \ref{sqdiagram}. A tuple is generated by a sensor and sent to the gateway the sensor is connected to. The callback function for handling an incoming tuple \textit{processTupleArrival()} is called once the tuple reaches the Fog device (gateway). In case the tuple needs to be routed to another Fog device, it is sent immediately without processing. Otherwise, if the application module on which the tuple needs to be executed is placed on the receiving Fog device, the tuple is submitted for execution. The function \textit{checkCloudletCompletion()} is called on the Fog device on completion of execution of the tuple. 


In addition to the basic tuple processing functionalities, simulated services available in \emph{iFogSim} are:

\begin{itemize}
\item \textbf{Power Monitoring Service:} Each Fog device (a FogDevice instance) is associated with a power model (e.g. PowerModelLinear). The \textit{executeTuple()} method in the \textit{FogDevice} class contains the tuple processing logic where the related power model is used to update the device power consumption based on the changes in the resource utilization.
\item \textbf{Resource Management Service:} iFogSim has two levels of resource management for applications --- Placement and Scheduling --- which are abstracted as separate policies to facilitate extension and customization.
\begin{enumerate}
	\item Application Placement: The placement policy determines how application modules are placed across Fog devices upon submission of application. The placement process can be driven by objectives such as minimizing end-to-end latency, network usage, operational cost, or energy consumption. The class \textit{ModulePlacement} is the abstract placement policy that needs to be extended for integrating new policies.	
	\item Application Scheduling: 
	Scheduling resources of the host Fog device to application modules forms the second level of resource management. The default resource scheduler equally divides a device's resources among all active application modules. The application scheduling policy can be customized by overriding the method \textit{updateAllocatedMips} inside the class \textit{FogDevice}. 
\end{enumerate}
\end{itemize}

\subsection{Built-in Module Placement Strategies}
iFogSim is packaged with two application module placement strategies --- \textit{cloud-only placement} and \textit{edge-ward placement}. 
\begin{enumerate}
\item \textbf{Cloud-only placement}: The \textit{cloud-only} placement strategy is based on the traditional cloud-based implementation of applications where all modules of an application run in data centers. The sense-process-actuate loop in such applications are implemented by having sensors transmiting sensed data to the cloud where it is processed and actuators are informed if action is required.

\item \textbf{Edge-ward placement}: Edge-ward placement strategy favours the deployment of application modules close to the edge of the network. However, devices close to the edge of the network --- like routers and access points --- may not be computationally powerful enough to host all operators of the application. In such a situation, the strategy iterates on Fog devices towards cloud and tries to place remaining operators on alternative devices. This strategy (shown in Algorithm \ref{algo:edgeward}) demonstrates the interplay between the Fog and the cloud by placing modules both near the network edge and the cloud.

\end{enumerate}

\begin{figure}[!t]
  \begin{algorithm}[H]
   \caption{Edge-ward module placement}
   \For( \emph{Across all paths}){$p \in PATHS$}
   {
      $placeList := \{ \} $\;     
      \For(leaf-to-root traversal){Fog device $d \in p$}
      {
         \For{module $w \in app$}
         {
            \If{all predecessors of $w$ are placed}
            {
               add $w$ to $placeList$ \;
            }
         }
         \For(){module $\theta \in placeList$}
         {
            \If{$\theta$ is already placed on device $f \in p$}
            {
            	   Merge $\theta$ with its upstream instance \;
            	   $f := $device holding merged instance \;
            	   \While{$CPU^{req}_{\theta} \geq CPU^{avail}_{f}$}
            	   {
            	      $f := parent(f)$ \;
            	   }
            	   Place $\theta$ on device $f$ \;
            }
            \ElseIf{$CPU^{req}_{\theta} \leq CPU^{avail}_{d}$}
            {
            	   Place $\theta$ on device $d$ \;
            }
         }
      }
   }
   \label{algo:edgeward}
  \end{algorithm}
\end{figure}

\subsection{Graphical User Interface}
\begin{figure*}[!htb]
\centering
\includegraphics[width = .9\textwidth]{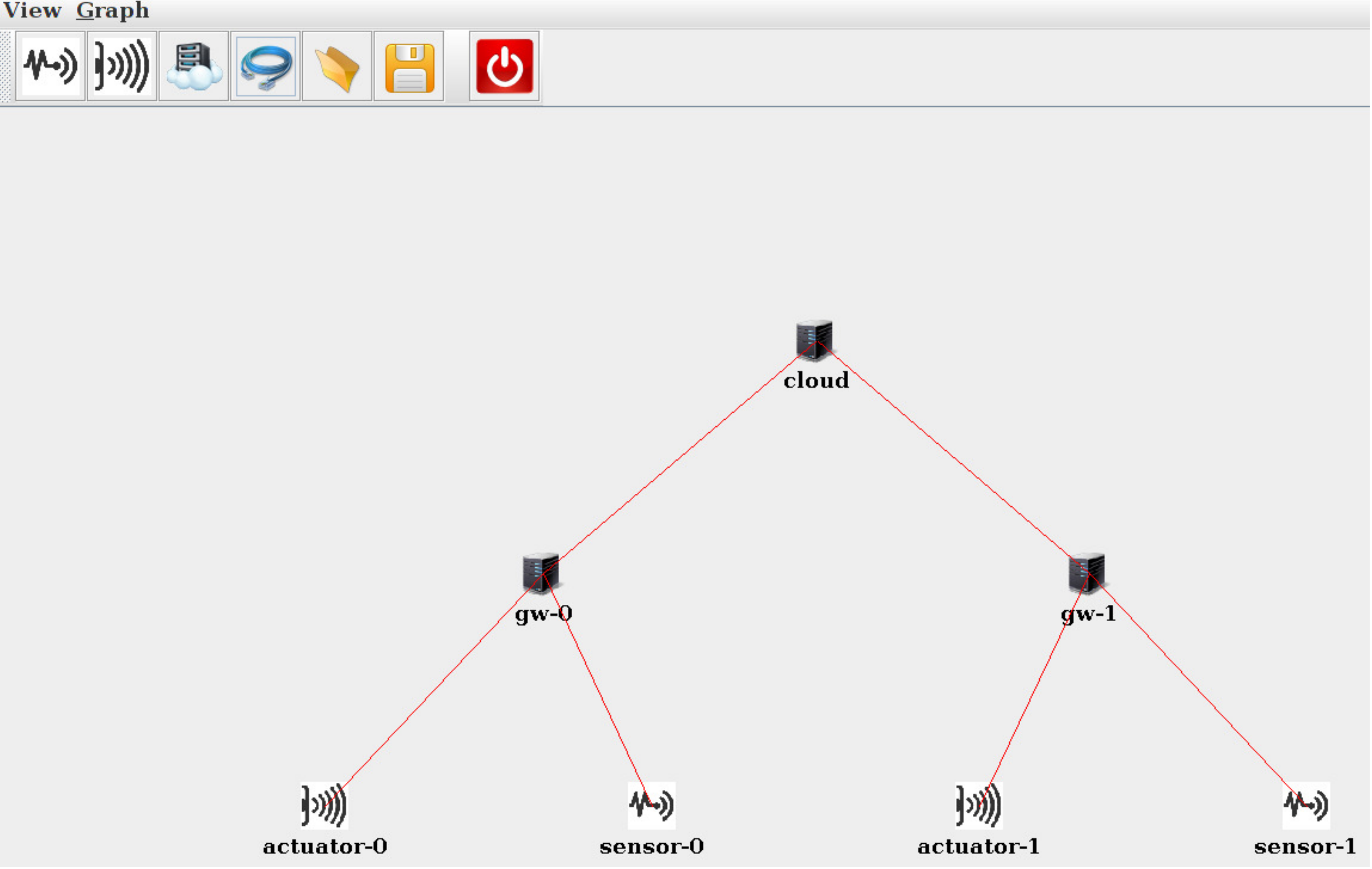}
\caption{iFogSim GUI for building network topology.}
\label{fig:topologyGUI}
\end{figure*}

\begin{figure*}[!htb]
\centering
\includegraphics[width = .9\textwidth]{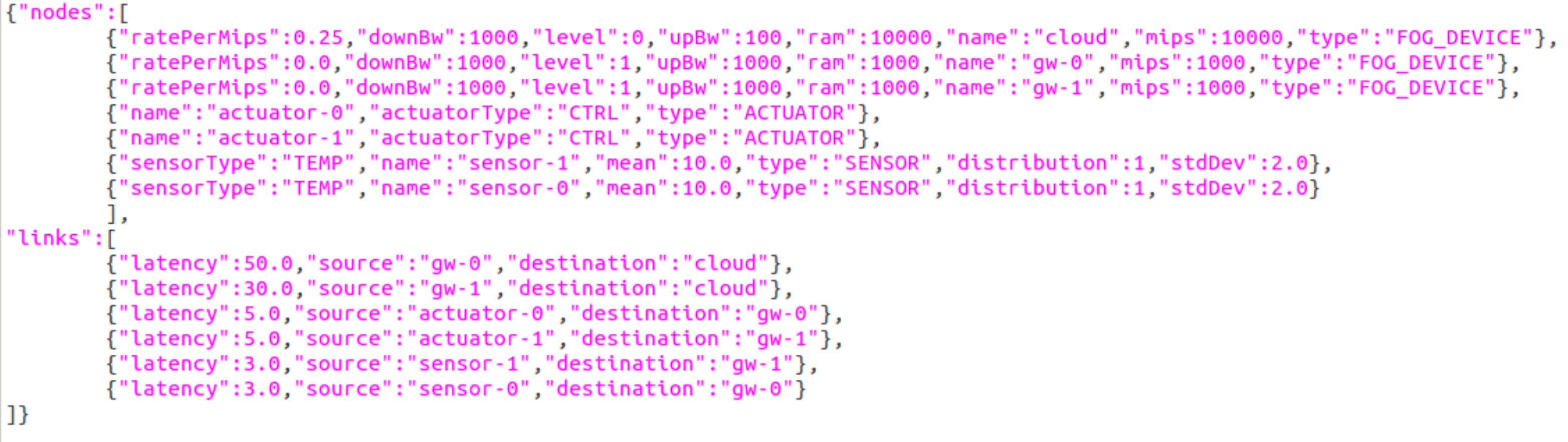}
\caption{Network topology JSON file.}
\label{fig:topologyJSON}
\end{figure*}

In order to facilitate description of the physical network topology, a GUI has been built over the iFogSim application logic. The GUI allows the user to draw physical elements such as Fog devices, sensors, actuators and connecting links. Defining characteristics of these entities can be fed to the topology using the GUI.
The drawn topologies can be saved and re-loaded by converting the topology to and from JSON file format. The physical topologies can be built both through GUI and programmatically through Java APIs. Figure \ref{fig:topologyGUI} shows a sample physical topology, which includes a sensor, a gateway, a Cloud virtual machine, and their connections. The JSON representation of this physical topology is shown in Figure \ref{fig:topologyJSON}.


\subsection{A Recipe for Simulating IoT Environments and Testing Resource Management Techniques}

This section lays down the high-level steps to simulate an IoT/Fog environment in iFogSim to analyze the performance of applications and resource management policies.
\begin{enumerate}
\item  First, physical entities need to be created and their capabilities and configurations specified. These include sensors, gateways and cloud virtual machines and the links that describe how these entities are connected. As mentioned earlier, this can be achieved either by using the GUI or programmatically using aforementioned iFogSim classes. 
To model the workload of the system, first we should set the tuple transmit rates of sensor using $transmitDistribution$ (an attribute in Sensor Class). Next step in modeling workload is identifying how much resources are required to process the tuples. For defining resource utilization including CPU and RAM, necessary variables are defined in the Tuple class.

\item Second, we need to model applications. As we stated earlier an application is modeled as a directed acyclic graph (DAG), and built via three classes of AppModule, AppEdge, and AppLoop. 
\item Finally, we need to define placement and scheduling policies that map application modules to Fog devices. The policies may consider range of criteria including end-to-end processing latency, throughput, cost, power consumption, and devices constraints. As we mentioned earlier, ModulePlacement and Controller classes are where the placement logic is implemented. 
\end{enumerate}

\section{Application Case Studies}
\label{sec:case_study}
In the this section, we provide two simulation case studies, namely a latency-sensitive online game and intelligent surveillance through distributed camera networks.  

\subsection{Case Study 1 --- A Latency-sensitive Online Game}
The latency-critical application in the case study is a human-vs-human game (\emph{EEG Tractor Beam Game}) \cite{zao2014augmented} that involves augmented brain-computer interaction.  In order to play the \emph{EEG Tractor Beam game}, each player needs to wear a MINDO-4S wireless EEG headset that is connected to his smartphone. The game runs as an Android application on a user's smartphone. The application performs real-time processing of the EEG signals sensed by the EEG headset and calculates the brain state of the user. 

\par On the application's display, the game shows all the players on a ring surrounding a target object. Each player can exert an attractive force onto the target in proportion to his level of concentration (estimated using a ratio of the average power spectral density in the EEG $\alpha$, $\beta$ and $\theta$ bands of the player). In order to win the game, a player should try to pull the target toward himself by exercising concentration while depriving other players of their chances to grab the target.

\par Real-time processing requires that the application be hosted very close to the source of data - preferably on the smartphone itself. However, such a deployment would not allow global coverage - which typically requires deploying the application in the cloud. Such a mix of conflicting objectives makes this application a typical use-case for Fog computing.

\par
\begin{figure}[!htb]
\centering
\includegraphics[width = .75\textwidth]{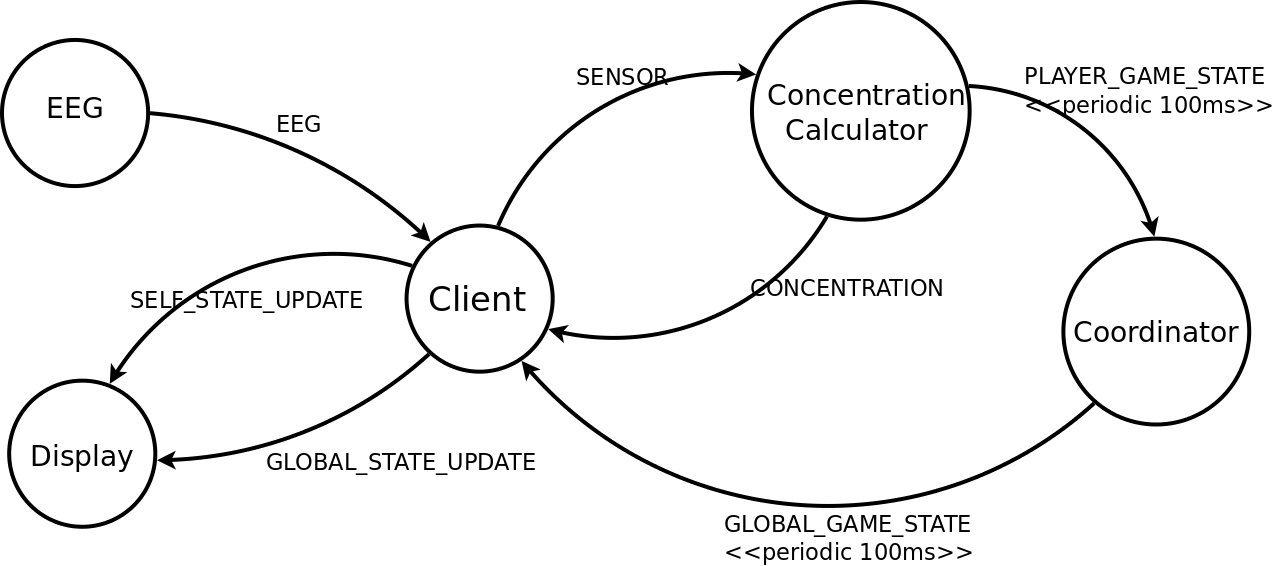}
\caption{Application model of the EEG game.}
\label{fig:vr_game_app}
\end{figure}

 \textbf{Application model}:
As illustrated in Figure \ref{fig:vr_game_app}, the application \emph{EEG Tractor Beam Game} consists of three major modules which perform processing - \textit{Client}, \textit{Concentration Calculator} and \textit{Coordinator}. The application modules are modeled in iFogSim using AppModule class. As depicted in  Figure \ref{fig:vr_game_app} there are data dependencies between modules, these dependences are modeled using AppEdge class in iFogSim. Finally, the control loop of interest for  EEG application is modeled in iFogsim using AppLoop class. The application is fed EEG signals by a sensor \textit{EEG} and an actuator \emph{DISPLAY} displays the current game-scene to the user. The functions of the above mentioned modules are as follows:
\begin{enumerate}
\item \textbf{Client}: Client module interfaces with the sensor and receives raw EEG signals. It checks the received signal values for any discrepancy and discards any seemingly inconsistent reading. If the sensed signal value is consistent, it sends the value to the \textit{Concentration Calculator} module to get the concentration level of the user from the signal. On receiving the concentration level, it displays it by sending the value to the actuator \textit{DISPLAY}.

\item \textbf{Concentration Calculator}: The concentration calculator module is responsible for determining the brain-state of the user from the sensed EEG signal values and calculating the concentration level. This module informs the \textit{Client} module about the measured concentration level so that the game state of the player on the display can be updated.

\item \textbf{Coordinator}: Coordinator works at the global level and coordinates the game between multiple players that may be present at geographically distributed locations. The \textit{Coordinator} continuously sends the current state of the game to the \textit{Client} module of all connected users.

\end{enumerate}

The properties of tuples (modeled using Tuple class) carried by edges between the modules in the application are described in Table \ref{table:vr_game_app_edges}.

\begin{table}
\caption{Description of inter-module edges in the EEG Tractor Game application}
\label{table:vr_game_app_edges}
\begin{center}
\begin{tabular}{ ||c|c|c|| } 
\hline
\textbf{TUPLE TYPE} & \textbf{CPU LENGTH} (MIPS) & \textbf{N/W LENGTH}\\
\hline \hline
EEG & 2000 (A) / 2500 (B) & 500\\
\hline
\_ SENSOR & 3500 & 500\\
\hline
PLAYER\_ GAME\_ STATE & 1000 & 1000\\
\hline
CONCENTRATION & 14 & 500\\
\hline
GLOBAL\_ GAME\_ STATE & 1000 & 1000\\
\hline
GLOBAL\_ STATE\_ UPDATE& 1000 & 500\\
\hline
SELF\_ STATE\_ UPDATE& 1000 & 500\\
\hline
\end{tabular}
\end{center}
\end{table}

\subsubsection{Physical Network}
\label{sec:pn}

\begin{table}
\caption{Configuration of Fog devices for EEG Tractor Game}
\label{vrdevice}
\begin{center}
\begin{tabular}{ ||c|c|c|c|c|c|c|c|| } 
\hline
\textbf{DEVICE TYPE} & \textbf{CPU GHz} & \textbf{RAM (GB)} & \textbf{POWER (W)}\\
\hline \hline
Cloud VM & 3.0 & 4 & 107.339(M) 83.433(I)\\
\hline
WiFi Gateway & 3.0 & 4 & 107.339(M) 83.433(I)\\
\hline
Smartphone & 1.6 & 1 & 87.53(M) 82.44(I)\\
\hline
ISP Gateway & 3.0 & 4 & 107.339(M) 83.433(I)\\
\hline
\end{tabular}
\end{center}
\end{table}

For the case study, we have considered a physical topology with 4 Fog devices. Table \ref{vrdevice} illustrates the configurations  \cite{guerout2013energy} of the different types of Fog devices used in the topology. Moreover, two different types of EEG headsets have been used --- each emitting tuples of different properties, as shown in Table \ref{tab:vrgamesensors}. The physical topology of the case study is modeled in iFogSim via FogDevice, Sensor, PhysicalTopology and Actuator classes.

\begin{table}
\caption{Configuration of sensors for EEG Tractor Game}
\label{tab:vrgamesensors}
\begin{center}
\begin{tabular}{ ||c|c|c|| } 
\hline
\textbf{HEADSET} & \textbf{TUPLE CPU LENGTH} &\textbf{Average Inter-arrival Time}\\
\hline \hline
A & 2000 Million Instructions & 10 milliseconds\\
\hline
B & 2500 Million Instructions & 5 milliseconds\\
\hline
\hline
\end{tabular}
\end{center}
\end{table}




\subsection{Case Study 2 --- Intelligent Surveillance through Distributed Camera Networks}
\label{sec:dcn}
Distributed system of cameras surveilling an area has garnered a lot of attention in recent years particularly by enabling a broad spectrum of interdisciplinary applications in areas of the likes of public safety and security, manufacturing, transportation, and healthcare. However, monitoring video streams from the system of cameras manually is not practical. Hence we need tools that automatically analyze data coming from cameras and summarize the results in a way that is beneficial to the end-user. The requirements of such a system have been listed down as follows.
\begin{itemize}
\item \textbf{Low-latency communication}: For effective object coverage, the Pan-tilt-zoom (PTZ) parameters of multiple cameras need to be tuned in real-time based on the captured image. This requires low latency communication between the cameras and the set of camera control strategies.
\item \textbf{Handling voluminous data}: Video cameras continuously send captured video frames for processing, which causes a huge traffic, especially when all cameras in a system are taken into account. It is necessary to handle such a large amount of data without burdening the network into a state of congestion.
\item \textbf{Heavy long-term processing}: The camera control strategy needs to be updated constantly so that it learns the optimal PTZ parameter calculation strategy. This requires analysis of the decisions taken by the control strategy over a long-period of time, which makes this analysis computationally intensive.
\end{itemize}
\par Centralized tools for analyzing camera-generated data are not desirable primarily because of the huge amount of data that needs to be sent to the central processing machine. This would not only lead to high latency in the system, but would also consume mote of available bandwidth. Hence, processing the video streams in a decentralized fashion is a more advisable method of analysis. 

The Intelligent Surveillance system aims at coordinating multiple cameras with different fields of view (FOVs) to surveil a given area. Coordination between cameras involves coordinated tuning of PTZ parameters so that the best view of the area can be obtained. Furthermore, the system alerts the user in case of irregular events --- which may demand attention of the security authorities.

\par The smart camera detects motion in its FOV and starts sending a video stream to the Intelligent Surveillance application. The application locates the moving object in the video stream sent and initiates tracking. Tracking of moving objects is done by constantly tuning the PTZ parameters of the cameras at that site so as to obtain the best view of all the tracked objects. Furthermore, in the event of detection of an event of interest, the application notifies the the system user and sends captured video streams to him through the Internet.

\par
\begin{figure}[!htb]
\centering
\includegraphics[width = .75\textwidth]{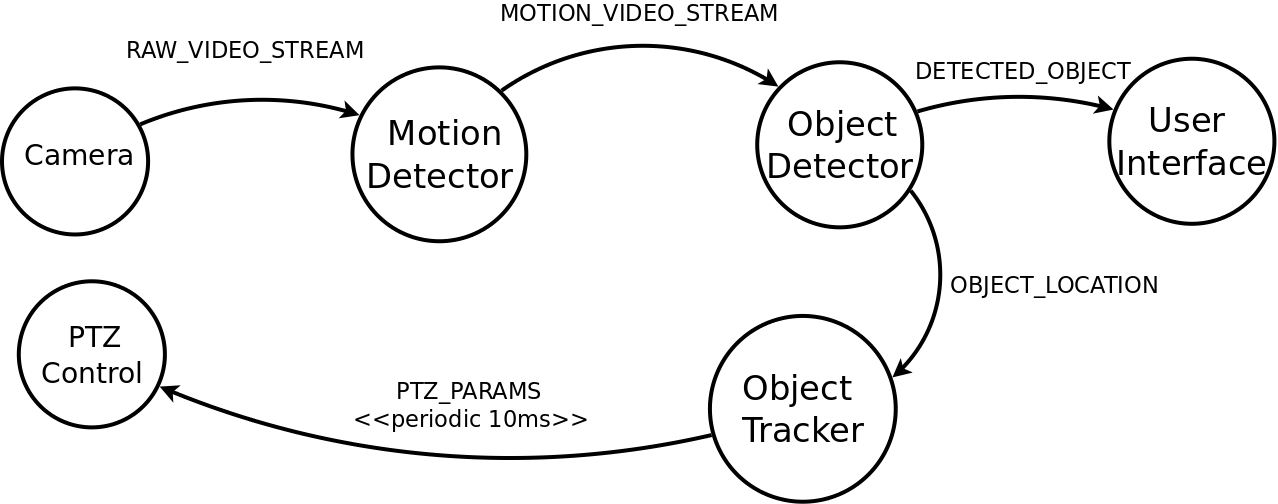}
\caption{Application Model of the Intelligent Surveillance Case Study.}
\label{fig:dcns_app}
\end{figure}

 \textbf{Application Model:}
As depicted in Figure \ref{fig:dcns_app}, the Intelligent Surveillance application consists of five major modules which perform processing --- \emph{Motion Detector}, \emph{Object Detector}, \emph{Object Tracker}, \emph{PTZ Control} and \emph{User Interface}. The application is fed live video streams by a number of CCTV cameras and the PTZ control in each camera continuously adjusts the PTZ parameters. The functions of the above mentioned modules are as follows:
\begin{enumerate}
\item \textbf{Motion Detection}: This module is embedded inside the smart cameras used in the case study. It continuously reads the raw video streams captured by the camera to find motion of an object. In the event of detection of motion in the camera's FOV, the video stream is forwarded upstream to the Object Detection module for further processing.
\item \textbf{Object Detection}: The Object Detection module receives video streams in which the smart cameras detect motion of an object. The module extracts the moving object from the video streams and compares them with previously discovered objects which are active in the area currently. In case the detected object has not been in the area before, tracking is activated for this object. In addition, it calculate the coordinates of the objects. 
\item \textbf{Object Tracker}: The Object Tracker module receives the last calculated coordinates of the currently tracked objects and calculates an optimal PTZ configuration of all the cameras covering the area so that the tracked objects can be captured in the most effective manner. This PTZ information is conveyed to the PTZ control of cameras periodically.
\item \textbf{PTZ Control}: This module runs on each smart camera and adjusts the physical camera to conform to the optimal PTZ parameters sent by the Object Tracker module. This module serves as the actuator of the system and is embedded in the smart cameras itself.
\item \textbf{User Interface}: The application presents a user interface by sending a fraction of the video streams containing each tracked object to the user's device. For this use-case, it requires such filtered video streams from the Object Detector module.
\end{enumerate}

Similar to the previous case studies The application modules , data dependencies, the control loop are modeled using AppModule, AppEdge, and AppLoop classes respectively. Finally, t of interest for  EEG application is modeled in iFogsim using  class. 

The properties of tuples carried by edges between the modules in the application are described in Table \ref{table:dcns_appedges}.

\begin{table}
\caption{Description of inter-module edges in the Intelligent Surveillance application}
\label{table:dcns_appedges}
\begin{center}
\begin{tabular}{ ||c|c|c|| } 
\hline
\textbf{Tuple Type} & \textbf{CPU Length} & \textbf{N/W Length}\\
\hline \hline
RAW\_ VIDEO\_ STREAM & 1000 & 20000\\
\hline
MOTION\_ VIDEO\_ STREAM & 2000 & 2000\\
\hline
DETECTED\_ OBJECT & 500 & 2000\\
\hline
OBJECT\_ LOCATION & 1000 & 100\\
\hline
PTZ\_ PARAMS & 100 & 100\\
\hline
\end{tabular}
\end{center}
\end{table}

\textbf{Physical Topology:} 
The physical topology for the second case study is similar to the first case study as described in Section \ref{sec:pn}.

%

Table \ref{table:dcns_sensors} shows the configuration of the sensors involved in the case study. Here, the cameras that recording live video feeds act as sensors and provide input data to the application. Similar to the previous case study, the physical topology is modeled in iFogSim via FogDevice, Sensor, PhysicalTopology and Actuator classes. Interested readers can check the examples in iFogSim package for more details on implementation of the case studies.

\begin{table}
\caption{Configuration of sensor for Intelligent Surveillance}
\label{table:dcns_sensors}
\begin{center}
\begin{tabular}{ ||c|c|c|| } 
\hline
\textbf{CPU Length} & \textbf{NW Length} &\textbf{Average Inter-arrival Time}\\
\hline \hline
1000 Million Instructions & 20000 bytes & 5 milliseconds\\
\hline
\end{tabular}
\end{center}
\end{table}

%
%
%
%
%

\section{Performance Evaluation}
\label{sec:pe}
In this section, we have simulated a Fog computing environment for the application case studies. Then, we evaluated efficiencies of the two placement strategies (i.e. cloud-only and edge-ward) in terms of latency, network usage, and energy consumption for each case study. Finally, we evaluated scalability of iFogSim in terms of RAM usage and execution time for different simulation scenarios.
\subsection{Evaluation of Case Study 1 --- A Latency-sensitive Online Game}
The simulation of this case study was carried out for a period of 3 hours and the various metrics reported by iFogSim were collected. The results of the simulation demonstrate how different input workloads and  placement strategies impacts the network usage and end-to-end latency.

Each headset is connected to a smartphone via Bluetooth communication link. Smartphones gain access to the Internet through WiFi gateways that are connected to the ISP Gateway. For the purpose of testing iFogSim's performance on varying topology sizes, we have varied the number of WiFi gateways keeping the number of smartphones connected to each gateway constant. Five configurations of physical topology have been simulated --- Config 1, Config 2, Config 3, Config 4, and Config 5 --- having 1, 2, 4, 8 and 16 WiFi gateways respectively, with each gateway connected to 4 smartphones playing the EEG Tractor Beam game. Performance of an application on the Fog depends on latencies of the links connecting the Fog devices. In the simulation topology, different kinds of devices have different latencies between them, which is listed down in Table \ref{table:vr_game_links}.

\begin{table}
\caption{Description of network links for EEG Tractor Game}
\label{table:vr_game_links}
\begin{center}
\begin{tabular}{ ||c|c|c|| } 
\hline
\textbf{Source} & \textbf{Destination} & \textbf{Latency} (in ms)\\
\hline \hline
EEG Headset & Smartphone & 6\\
\hline
Smartphone & WiFi Gateway & 2\\
\hline
WiFi Gateway & ISP Gateway & 4\\
\hline
ISP Gateway & Cloud DC & 100\\
\hline
\end{tabular}
\end{center}
\end{table}

\subsubsection{Average Latency of Control Loop.}
\begin{figure}[!htb]
\centering
\includegraphics[width = 0.8 \textwidth]{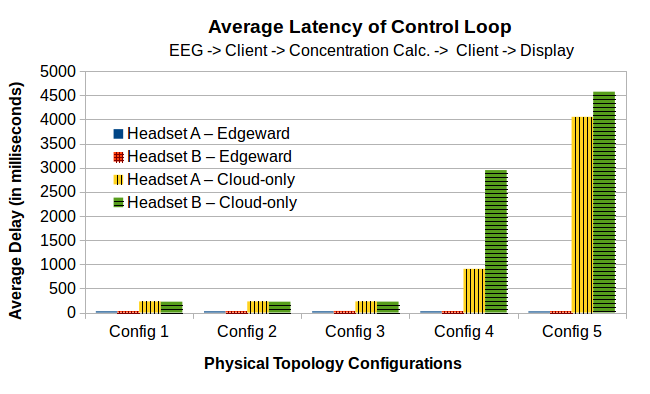}
\caption{Average delay of control loop.}
\label{fig:vr_game_delay}
\end{figure}

The most important control loop in the EEG Tractor game application --- in terms of latency of response --- is the loop responsible for transforming the brain-state of the user into his game-state on the smartphone's display. This requires real-time communication between the smartphone and the device hosting the brain-state classification module along with efficient processing on the classification module. Lag in this loop will severely mar user experience as it affects entities that the user directly interacts with. Figure \ref{fig:vr_game_delay} illustrates the average delay in execution of this control loop.
Figure \ref{fig:vr_game_delay} shows that control loop execution delay dramatically decreases for Edge-ward placement strategy where Fog devices are utilized for processing. This reduction is even more pronounced when topology sizes and tuple emission rate increases (Headset B).  

\subsubsection{Network Usage.}
\begin{figure}[!htb]
\centering
\includegraphics[width = .8 \textwidth]{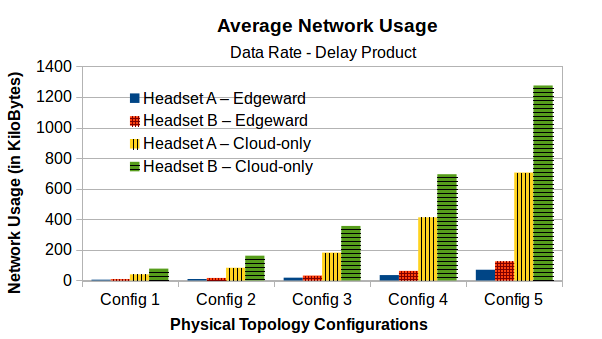}
\caption{Comparison of network usage.}
\label{fig:vr_game_us_usage}
\end{figure}

Figure \ref{fig:vr_game_us_usage} shows the network usage of the EEG Tractor Beam game application. Increase in the number of devices connected to the application significantly increases the load on the network   where only cloud resources used. As Figure \ref{fig:vr_game_us_usage} shows, when Fog devices are considered, the network usage considerably decreased. 

This result can also be interpreted as a demonstration of scalability of Fog-based applications. Uncontrolled growth of network usage in case of cloud-based execution can lead to network congestion and lead to further degradation of the application's performance. Such situations can be better avoided if Fog-based deployment is adopted.

\subsubsection{Energy Consumption.}
\begin{figure}[!htb]
\centering
\includegraphics[width = 0.8\textwidth]{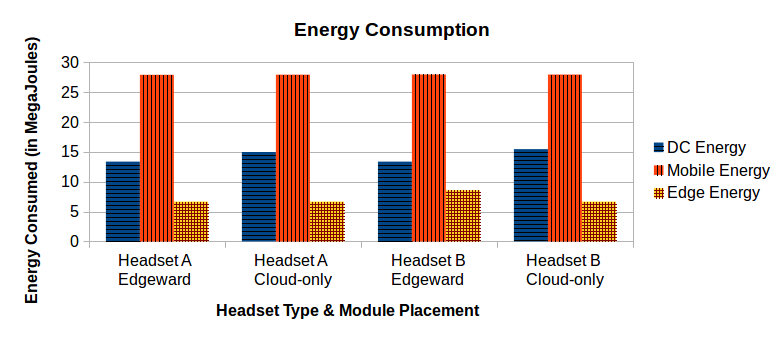}
\caption{Energy consumption of devices in cloud and Fog execution.}
\label{fig:vr_game_energy}
\end{figure}

Figure \ref{fig:vr_game_energy} portrays the energy consumed by different classes of devices in the simulation. 
As Figure \ref{fig:vr_game_energy} shows, utilizing Fog devices in Edge-ward placement strategy reduces energy consumption of cloud data centers while slightly increases energy consumption of edge devices.

\subsection{Evaluation of Case Study 2 --- Intelligent Surveillance through Distributed Camera Networks}
For demonstrating the flexibility of iFogSim, the Intelligent Surveillance application has been evaluated on a number of physical infrastructure configurations. The number of surveilled areas has been varied from 1 to 16. Note that each surveilled area has four smart cameras monitoring the area. These cameras are connected to an area gateway that manages the activity in that surveilled area. In the simulated topologies, each surveilled area has 4 smart cameras connected to an area gateway, which is responsible for providing Internet access to them. The number of surveilled areas is varied across physical topology configurations Config 1, Config 2, Config 3, Config 4 and Config 5, having 1, 2, 4, 8 and 16 surveilled areas respectively. The network latencies between devices are listed in Table \ref{table:dcns_links}.

\begin{table}
\caption{Description of network links in the physical topology for Intelligent Surveillance}
\label{table:dcns_links}
\begin{center}
\begin{tabular}{ ||c|c|c|| } 
\hline
\textbf{Source} & \textbf{Destination} & \textbf{Latency} (ms)\\
\hline \hline
Camera & Area Switch & 2\\
\hline
Area GW & ISP Gateway & 2\\
\hline
ISP Gateway & Cloud DC & 100\\
\hline
\end{tabular}
\end{center}
\end{table}

Based on the aforementioned configurations of entities, a physical topology is designed. The topology has the cloud data center at the apex and smart cameras at the edge of the network. Smart cameras are fed live video streams in the form of tuples for performing motion detection and the PTZ control of the camera has been modeled as an actuator. Similarly, two placement strategies , namely \textit{cloud-only} and \textit{Edge-ward} are used for placing application modules on the physical network. In case of \textit{cloud-only} placement, all operators in the application are placed on the cloud data center except the \textit{Motion Detector} module, which is bound to the smart cameras. However, in the \textit{Edge-ward} placement, the \textit{Object Detector} and \textit{Object Tracker} modules are pushed to WiFi gateways connecting the cameras in a surveilled area to the Internet. The simulation of this case study was carried out for a period of 1000 seconds. 

\subsubsection{Average Latency of Control Loop.}

\begin{figure}[!htb]
\centering
\includegraphics[width = 0.8\textwidth]{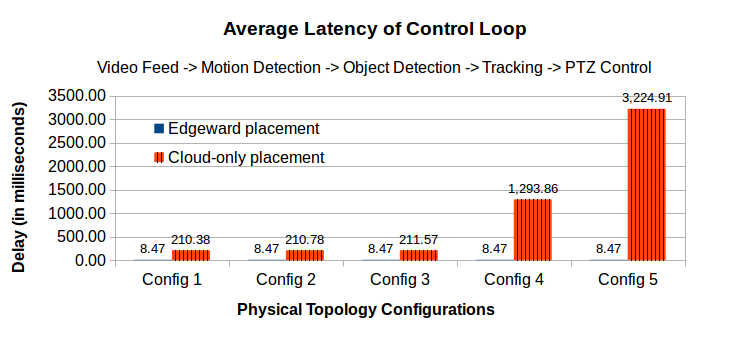}
\caption{Average delay of control loop.}
\label{fig:dcns_delay}
\end{figure}

Figure \ref{fig:dcns_delay} demonstrates the average processing latency of sensing-actuation control loop.

In the case of \textit{cloud-only} placement strategy, as Figure \ref{fig:dcns_delay} shows,  cloud data centers turned to a bottleneck in execution of the modules which caused a notably significant increase in latency.  On the other hand, the \textit{Edge-ward} placement succeeds in maintaining low latency, as it places the modules critical to the control loop close to the network edge.

\subsubsection{Network Usage.}

\begin{figure}[!htb]
\centering
\includegraphics[width = 0.8\textwidth]{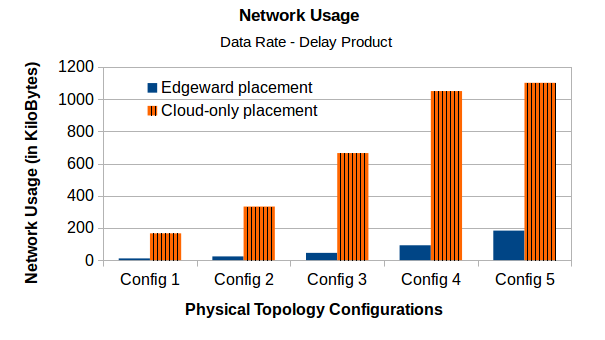}
\caption{Comparison of network usage.}
\label{fig:dcns_nw_usage}
\end{figure}

Figure \ref{fig:dcns_nw_usage} shows the network usage of the Intelligent Surveillance application for the placement strategies. As number of devices connected to the application increases, the load on the network increases significantly in the case of cloud-only deployment in contrast to edge-ward deployment.

This observation can be attributed to the fact that in the \textit{Fog-based} execution, most of the data-intensive communication takes place through low-latency links. Hence, modules like Object Detector and Object Tracker are placed on the edge devices, which substantially decrease the volume of data sent to a centralized cloud data center.
\subsubsection{Energy Consumption.}
\begin{figure}[!htb]
\centering
\includegraphics[width = 0.8\textwidth]{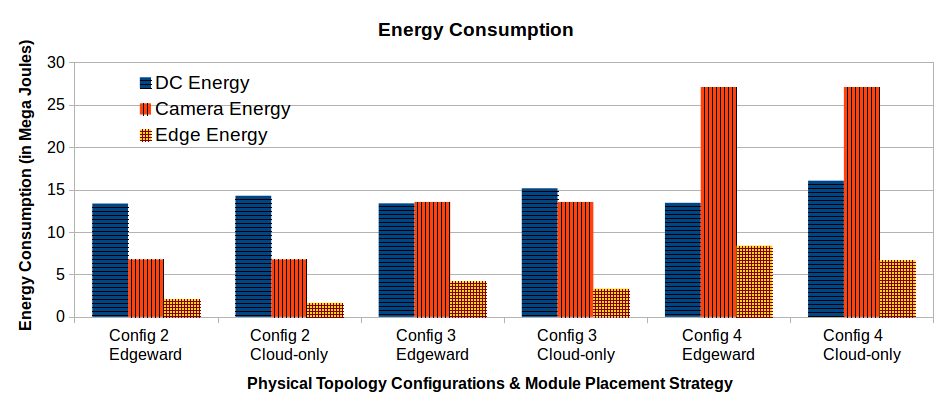}
\caption{Energy consumption of devices in Fog execution.}
\label{fig:dcns_energy}
\end{figure}

Figure \ref{fig:dcns_energy} shows the energy consumed by different category of devices in the simulation. Deployment of application on Fog devices has been compared to deployment only on the cloud data centers. Cameras perform motion detection in the captured video frames, which drains out a large amount of power. Therefore, as Figure \ref{fig:dcns_energy} shows, when areas under surveillance increases, energy consumption in these devices increase too. Furthermore, like the previous case study, the energy consumption in the cloud data center decreases when operators are pushed to Fog devices. 

\subsection{iFogSim Execution Footprint Analysis}
The scalability of a simulator depends on it resource usage (in particular RAM) and the time it takes for simulation to execute. The following metrics have been collected for various sizes of simulation and reported in this section. We have used the configuration of the first case study and the Headset-B emits tuples at twice the rate of Headset-A.

\subsubsection{RAM Usage.}

The \textit{massif} heap profiler available in  \textit{Valgrind} tool suite \cite{seward2004valgrind} was used to measure the heap allocations during simulations of various topology sizes and input workloads. 
\begin{figure}[!htb]
\centering
\includegraphics[width = 0.8 \textwidth]{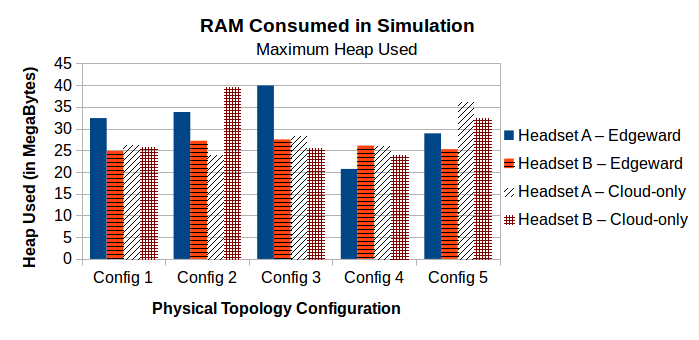}
\caption{RAM usage of simulation for varying sizes of topology and input workload.}
\label{fig:vr_game_ram}
\end{figure}
As depicted in Figure \ref{fig:vr_game_ram}, the heap allocation does not increase considerably with increasing workload and physical topology size.  The Figure shows that iFogSim scales with minimal memory overhead when the number of sensors (smart phones) and gateways increases from 4 to 64 and 1 to 16.

\subsubsection{Simulation Time.}

\begin{figure}[!htb]
\centering
\includegraphics[width = 0.8 \textwidth]{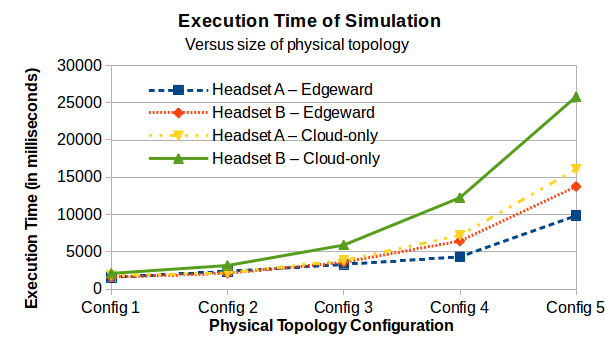}
\caption{Execution time of simulation for varying sizes of topology and input workload.}
\label{fig:vr_game_time}
\end{figure}

Simulation time for various topologies and input workloads was measured and reported in Figure \ref{fig:vr_game_time}. 

Figure \ref{fig:vr_game_time} shows that the execution time  increases when number of devices and transmission rate increases. However, the increase in simulation is almost linear and consequently simulation  can be executed in an acceptable time (25 seconds) even if a considerable number of gateways is added.

%
%
%
%

\section{Related Work}

In IoT environments \cite{gubbi2013internet}, in order to enable real-time decision making \cite{stojmenovic2015overview}, distributed stream-processing systems have to push query operators to nodes located closer to the source of data. To this end, Cisco introduced Fog computing \cite{bonomi2014fog}.  Cisco's offering for Fog computing, known as IOx \cite{ CiscoIo:online}, is a combination of  networking operating system, IOS and the most popular open source Operating System, Linux. Ruggedized routers running Cisco IOx make compute and storage available to applications hosted in a Guest Operating System running on a hypervisor alongside the IOS virtual machine. Cisco provides an app-store which allows users to download applications to IOx devices and an app-management console for controlling and monitoring the performance of an application. Using device abstractions provided by Cisco IOx APIs, applications running on the Fog can communicate with IoT devices that use verities of protocols. In iFogSim also we have considered the  similar concepts of apps and devices to model a Fog environment and controllers which act similar to app-management consoles in Cisco environments.

The FIT IoT-LAB \cite{adjih2015fit} is a testbed equipped with thousands of wireless nodes located in six different sites across France.  It allows users to evaluate and test their novel ideas ranging from low level protocols to advanced Analytic and services in a very large scale wireless IoT environment.  Major services offered by IoT-LAB include: 1) Remote access to sensors and gateways: the testbed provides users with APIs to flash any firmware, design, build, and compile applications; 2) Access to the serial ports of reserved IoT devices; 3) Internet access for nodes with end-to-end IP connection using IPv6 and 6LoWPAN; 4) Power consumption monitoring per device; 5) and robots to test and improve real-time decision making in IoT context. 

SmartSantander \cite{sanchez2014smartsantander} is a project of the Future Internet Research and Experimentation initiative of the European Commission. It uniquely offers a city-scale experimental research facility with support of services of a smart city. The testbed comprises a large number of Internet of Things devices deployed in several urban locations mainly in Santander city.  SmartSantander has conceived a 3-tiered architecture as follows: 1) IoT nodes: Responsible for sensing the environment parameter such as temperature and noise; 2) Repeaters: These nodes are placed between sensors and gateways, in order to behave as forwarding nodes; and 3) Gateways:  IoT nodes and repeaters are configured to send all captured data via  802.15.4 protocol to gateways.

Simulators are essential tools during the design of IoT systems. A real IoT testbed ---which can be built using Cisco solutions or combination of open-source solutions such as FIT IoT Lab --- although  desirable, in many cases is too costly and does not provide repeatable and controllable environment. Therefore, simulation can be considered as cost-effective first step before real experimentation to eliminate ineffective policies and strategies. 

WSNet~\cite{chelius2006wsnet} is an event-driven simulator for wireless networks which can as well be used for IoT. It is capable of simulating nodes with different energy sources, mobility models, radio interfaces, applications, and routing protocols. Environment simulation is also supported by WSNet, in fact it offers the opportunity for modelling and simulation of physical phenomena (e.g. fire) and physical measures (e.g. temperature, humidity). These values (e.g. temperature) can be observed by the nodes, and can also impact nodes.

SimpleIoTSimulator \cite{SimpleIoTSimulator} is a commercial simulator for creating IoT environments consisting of many sensors and gateway. SimpleIoTSimulator supports common IoT protocols including CoAP and MQTT as a publish/subscribe based protocol. SimpleIoTSimulator objective is enabling IoT platform and gateway vendors to improve product quality with the focus on communication protocols. Our simulators also models publish/subscribe based protocols, however our focus is on the analysis of application design and resource management policies. In addition, the SimpleIoTSimulator dose not model Fog environments where services can be deployed both on edge and cloud resources.

Since traditional wireless sensor networks and IoT simulators do not focus on modeling of large scale deployments, Giacomo et al.~\cite{brambilla2014simulation} proposed a simulation methodology for IoT systems with a large number of interconnected devices. It is designed to study low-level networking aspects. In summary, the main advantages of their approach are 1) simulation of IoT systems with geographically distributed devices; 2) simulation of are IoT devices with multiple network interfaces and protocols, as well as different mobility, network, and energy consumption models.

The OASIS standard Devices Profile for Web Services (DPWS) aims at enabling the deployment of web services on constrained devices. To accelerate the development of DPWS enabled applications, Han et al. proposed DPWSim \cite{6803226}, a simulation toolkit that allows developers to design, develop, and test service-based IoT applications using the DPWS technology without the presence of physical sensors and gateways. 

CloudSim is developed as an extensible cloud simulation toolkit that enables modeling and simulation of cloud systems and application provisioning environments~\cite{cloudsim}. This toolkit provides both system and behaviour modeling of cloud computing components such as virtual machines (VMs), data centers, and users. However, CloudSim and other cloud environment simulators such as GloudSim~\cite{diGloudsim:2014}, DCSim~\cite{TigheDCSim:2012},  and GroudSim~\cite{OstermannGroudSim:2010}  do not model IoT devices and stream processing applications.

In summary, although there few simulators that model IoT environments, iFogSim is uniquely designed and implemented to model Fog environment along with IoT and cloud. This enables innovation and performance evaluation of resource management policies for IoT applications such as real-time stream processing in a comprehensive end-to-end environment.  

\section{Conclusions and Future Directions}
\label{sec:conclusion}
Fog  and Edge computing are emerging as an attractive solutions to the problem of data processing in the Internet of Things. Rather than outsourcing all operations to cloud, they also utilize devices on the edge of the network that have more processing power than the end devices and are closer to sensors, thus reducing latency and network congestion. In this paper, we introduced \emph{iFogSim} to model and simulate IoT, Fog, and Edge computing environments. In particular, \emph{iFogSim} allows investigation and comparison of resource management techniques based on QoS criteria such as latency under different workloads (tuple size and transmit rate).  We described two case studies and demonstrated effectiveness of \emph{iFogSim} for evaluating resource management techniques including cloud-only application module placement and a techniques that pushes  applications towards edge devices when enough resources are available. Moreover, scalability of simulation is verified. Our experiment results demonstrated that \emph{iFogSim} is capable of supporting simulations on the scale expected in the context of IoT. We also believe that the availability of our simulator will energize rapid development of innovative resource management policies in the areas of IoT and Fog computing with end-to-end modeling and simulation.

There are a number of future directions that can enhance iFogSim capabilities and resource management strategies in the context of IoT:

\begin{itemize}
	
	\item \textbf{Power-Aware resource management policies:} One of the biggest challenges that most of Fog computing solutions face is how to get extra bit of battery life for Fog devices. To this end, future studies can look into new policies that dynamically and based on the  battery life of devices migrate the operators. Questions such as which operator to migrate, when to migrate, and where to migrate need to be addressed in by these policies.   
	
	\item \textbf{Priority-aware resource management strategies for multi-tenant environments:} Looking into scheduling policies for an environment where multiple application instances (DAGs of operators) share the same pool of resources and are assigned different Service Level Objectives (SLO) is another promising research direction.
		
	\item \textbf{Modeling failures of Fog devices:} Future research can focus on extracting failure models for the dominant failures in IoT and Fog devices. The developed models can be used to evaluate and compare reliability-aware scheduling and recovery policies for a wide range of applications.
	
	\item \textbf{Dynamic priority and SLA aware flow placement and resource scheduling (joint Edge-Network resource optimization):} In IoT environments, heterogeneous network and sensing resources have to be often shared with multiple applications or services with different and dynamic quality of service requirements. Therefore, the joint Edge-Network resource scheduling problem is another problem that we are going to investigate.
	\item \textbf{Modeling and comparison different virtualization techniques of IoT environments:} Future research studies can also consider and compare the performance of full virtualization, para-virtualization (as instances of hardware-level virtualization), and operating system level virtualization such as containers. 

\end{itemize}
\section*{Software Availability}
iFogSim  is available for download from the CLOUDS Lab website: \url{ http://www.cloudbus.org/cloudsim}.
%
%
%
\section*{Acknowledgements}
This work is: (i) supported by Melbourne-Chindia Cloud Computing Research Network and (ii) initiated as part of the first author's visit to CLOUDS Lab at the University of Melbourne. It is also partially supported by the ARC Future Fellowship.

\bibliographystyle{wileyj}

\bibliography{IEEEabrv,references}

\begin{thebibliography}{10}
\providecommand{\url}[1]{\texttt{#1}}
\providecommand{\urlprefix}{URL }
\expandafter\ifx\csname urlstyle\endcsname\relax
  \providecommand{\doi}[1]{doi:\discretionary{}{}{}#1}\else
  \providecommand{\doi}{doi:\discretionary{}{}{}\begingroup
  \urlstyle{rm}\Url}\fi

\bibitem{chang2015middleware}
Chang C, Srirama SN, Mass J. A middleware for discovering proximity-based
  service-oriented industrial internet of things. \emph{Services Computing
  (SCC), 2015 IEEE International Conference on}, IEEE, 2015; 130--137.

\bibitem{bonomi2014fog}
Bonomi F, Milito R, Natarajan P, Zhu J. Fog computing: A platform for internet
  of things and analytics. \emph{Big Data and Internet of Things: A Roadmap for
  Smart Environments}. Springer, 2014; 169--186.

\bibitem{giang2015developing}
Giang NK, Blackstock M, Lea R, Leung V. Developing iot applications in the fog:
  a distributed dataflow approach. \emph{Internet of Things (IOT), 2015 5th
  International Conference on the}, IEEE, 2015; 155--162.

\bibitem{calheiros2011cloudsim}
Calheiros RN, Ranjan R, Beloglazov A, De~Rose CA, Buyya R. Cloudsim: a toolkit
  for modeling and simulation of cloud computing environments and evaluation of
  resource provisioning algorithms. \emph{Software: Practice and Experience}
  2011; \textbf{41}(1):23--50.

\bibitem{zao2014augmented}
Zao JK, Gan TT, You CK, Rodríguez~Méndez SJ, Chung CE, Te~Wang Y, Mullen T,
  Jung TP. Augmented brain computer interaction based on fog computing and
  linked data. \emph{Intelligent Environments (IE), 2014 International
  Conference on}, IEEE, 2014; 374--377.

\bibitem{guerout2013energy}
Gu{\'e}rout T, Monteil T, Da~Costa G, Calheiros RN, Buyya R, Alexandru M.
  Energy-aware simulation with dvfs. \emph{Simulation Modelling Practice and
  Theory}  2013; \textbf{39}:76--91.

\bibitem{seward2004valgrind}
Seward J, Nethercote N, Fitzhardinge J. Valgrind, an open-source memory
  debugger for x86-gnu/linux. \emph{URL: http://www. ukuug.
  org/events/linux2002/papers/html/valgrind}  2004; .

\bibitem{gubbi2013internet}
Gubbi J, Buyya R, Marusic S, Palaniswami M. Internet of things (iot): A vision,
  architectural elements, and future directions. \emph{Future Generation
  Computer Systems}  2013; \textbf{29}(7):1645--1660.

\bibitem{stojmenovic2015overview}
Stojmenovic I, Wen S, Huang X, Luan H. An overview of fog computing and its
  security issues. \emph{Concurrency and Computation: Practice and Experience}
  2015; .

\bibitem{CiscoIo:online}
Iox overview.
  \url{https://developer.cisco.com/site/iox/documents/developer-guide/?ref=overview}.
  (Accessed on 05/11/2016).

\bibitem{adjih2015fit}
Adjih C, Baccelli E, Fleury E, Harter G, Mitton N, Noel T, Pissard-Gibollet R,
  Saint-Marcel F, Schreiner G, Vandaele J, \emph{et~al.}. Fit iot-lab: A large
  scale open experimental iot testbed. \emph{Proceedings of the 2nd IEEE World
  Forum on Internet of Things (WF-IoT)}, 2015.

\bibitem{sanchez2014smartsantander}
Sanchez L, Mu{\~n}oz L, Galache JA, Sotres P, Santana JR, Gutierrez V, Ramdhany
  R, Gluhak A, Krco S, Theodoridis E, \emph{et~al.}. Smartsantander: Iot
  experimentation over a smart city testbed. \emph{Computer Networks}  2014;
  \textbf{61}:217--238.

\bibitem{chelius2006wsnet}
Chelius G, Fraboulet A, Fleury E. Wsnet: a modular event-driven wireless
  network simulator  2006; .

\bibitem{SimpleIoTSimulator}
Simpleiotsimulator: The internetofthings simulator.
  \url{http://www.smplsft.com/SimpleIoTSimulator.html}. (Accessed on
  05/11/2016).

\bibitem{brambilla2014simulation}
Brambilla G, Picone M, Cirani S, Amoretti M, Zanichelli F. A simulation
  platform for large-scale internet of things scenarios in urban environments.
  \emph{Proceedings of the First International Conference on IoT in Urban
  Space}, ICST (Institute for Computer Sciences, Social-Informatics and
  Telecommunications Engineering), 2014; 50--55.

\bibitem{6803226}
Han SN, Lee GM, Crespi N, Luong NV, Heo K, Brut M, Gatellier P. Dpwsim: A
  simulation toolkit for iot applications using devices profile for web
  services. \emph{Internet of Things (WF-IoT), 2014 IEEE World Forum on}, 2014;
  544--547.

\bibitem{cloudsim}
Calheiros RN, Ranjan R, Beloglazov A, De~Rose CA, Buyya R. {CloudSim}: a
  toolkit for modeling and simulation of cloud computing environments and
  evaluation of resource provisioning algorithms. \emph{Software: Practice and
  Experience}  2011; \textbf{41}(1):23--50.

\bibitem{diGloudsim:2014}
Di S, Cappello F. {GloudSim:} google trace based cloud simulator with virtual
  machines. \emph{Software: Practice and Experience}  2014; .

\bibitem{TigheDCSim:2012}
Tighe M, Keller G, Bauer M, Lutfiyya H. {DCSim:} a data centre simulation tool
  for evaluating dynamic virtualized resource management. \emph{Proceedings of
  the 2012 8th international conference on Network and service management
  (cnsm) and 2012 workshop on systems virtualiztion management (svm)}, 2012;
  385--392.

\bibitem{OstermannGroudSim:2010}
Ostermann S, Plankensteiner K, Prodan R, Fahringer T. {GroudSim:} an
  event-based simulation framework for computational grids and clouds.
  \emph{Proceedings of the 2010 Conference on Parallel Processing}, Euro-Par
  2010, Springer-Verlag: Berlin, Heidelberg, 2011; 305--313.

\end{thebibliography}
\end{document}